\documentclass[preprint,12pt]{elsarticle}

\usepackage{amssymb,amsmath,amsthm,bm}
\usepackage[figuresright]{rotating}

\newcommand{\bs}{\boldsymbol}
\newcommand{\rme}{\mathrm{e}}
\newcommand{\rmi}{\mathrm{i}}

\newcommand{\tp}{\tilde{p}_3}
\newcommand{\tpm}{|\tilde{p}_3|}
\newcommand{\be}{\begin{equation}}
\newcommand{\ee}{\end{equation}}

\usepackage{amsfonts}
\usepackage{amssymb}
\usepackage{amsmath}
\usepackage{pst-plot}
\usepackage{feynmp}
\usepackage{graphicx}
\usepackage{wrapfig}

\biboptions{sort&compress}

\begin{document}

\begin{frontmatter}



\title{Spin light of relativistic electrons in neutrino fluxes}


\author[1]{Ilya A.~Balantsev}
\ead{balantsev@physics.msu.ru}

\author[1,2]{Alexander I. Studenikin}
\ead{studenik@srd.sinp.msu.ru}

\address[1]{Department of Theoretical Physics, Faculty of Physics, Lomonosov Moscow State University, 119991 Moscow, Russia}
\address[2]{Joint Institute for Nuclear Research, 141980 Dubna , Moscow Region, Russia}

\begin{abstract}

A new mechanism of electromagnetic radiation by electrons under the influence of a dense neutrino flux, termed ``the spin light of electron'' in neutrino flux ($SLe_\nu$), is considered. It is shown that in the case when  electrons are moving against the neutrino flux with relativistic energy there is a reasonable increase of the efficiency of the energy transfer from the neutrino flux to the electromagnetic radiation by the $SLe_\nu$ mechanism. The proposed radiation process is applied to an astrophysical environment with characteristics peculiar to supernovae. It is shown that a reasonable portion of energy of the neutrino flux can be transferred by the $SLe_\nu$ to gamma-rays .

\end{abstract}


\end{frontmatter}





\biboptions{comma, square, sort&compress}




\section{Introduction}
\label{Introduction}

For a period of about a decade neutrino electromagnetic properties (see \cite{Giunti:2014,Broggini:2012df,Giunti:2008ve} for a review) and neutrino electromagnetic interactions in dense matter have been under the focus of studies performed at the neutrino theory group at the Moscow State University. Within these studies, in particular, a new mechanism of electromagnetic radiation that can be emitted by a neutrino with nonzero magnetic moment propagating in dense matter was proposed and termed the spin
light of neutrino in matter ($SL\nu$) \cite{Lobanov:2002ur}. The quantum theory of the $SL\nu$ was first revealed in our studies \cite{Studenikin:2004dx,Grigorev:2005sw} ( see also \cite{Lobanov:2005plb}) within implication of  the so called ``method of wave equations exact solutions" that implies use of exact solutions of modified Dirac equations that contain the corresponding effective potentials accounting for the matter influence on neutrinos \cite{Studenikin:2004dx,Grigorev:2005sw,Studenikin:2005bq,Studenikin:2007zza,Studenikin:2008qk,Balantsev:2011jp,Balantsev:2013pan,StudenikinTokarev:2014npb}.

In this short note we continue studies of a new possible realization of the spin light mechanism of electromagnetic radiation in a dense environment that was termed the ``spin light of electron" ($SLe_\nu$) in a dense neutrino flux \cite{Balantsev:2014sle}. This phenomenon is a new mechanism of electromagnetic radiation that can be emitted by an electron in a dense flux of ultra-relativistic neutrinos. This new scheme of the spin light provides a possibility to avoid two suppression factors in the radiation rate and power peculiar for the $SL\nu$: 1) a suppression due to smallness of a neutrino magnetic moment and 2) a suppression due to effects of the background plasma.

One might expect that the predicted $SLe_\nu$ mechanism can have visible consequences for different astrophysical settings, for stellar core-collapse and supernova explosion phenomenology in particular. However, as it was shown in \cite{Balantsev:2014sle}, the  $SLe_\nu$ in a dense neutrino flux in the case of emitting electrons are at rest cannot provide important consequences for the energy balance in a supernova process. This is because in case of nonmoving electrons the emitted photon energy in the $SLe_\nu$ process is very small as well as  the rate of the process is also very small. Here below we consider ``strengthening" the $SLe_\nu$ in case of the relativistic motion of the emitting electrons. It is shown that the $SLe_\nu$ rate and power are increased by many orders of magnitude in respect to the case of the $SLe_\nu$ by electrons at rest. Also the emitted photon energies are shifted up to gamma-rays.

\section{Modified Dirac equation}

We consider a beam of electrons moving towards the neutrino flux
composed of three flavors $\nu_e$, $\nu_{\mu}$ and $\nu_{\tau}$
with number densities $n_{i}$ (in the laboratory rest frame) moving in the same direction.
Following discussion of \cite{Balantsev:2014sle}, we introduce the average value $n$ of the neutrino number density and the parameter $\delta_e$,
\be \label{n_delta}
n=\frac{n_e+n_\mu+n_\tau}{3},\qquad \delta_{e}=\frac{n_\mu+n_\tau-n_e}{n},
\ee
and obtain the modified Dirac equation for an electron in the neutrino flux,
\be\label{modified Dirac equation}
\{\gamma_{\mu}p^{\mu}+\gamma_{\mu}\frac{c+\delta_e\gamma^5}{2}f^{\mu}-m\}\Psi(x)=0,
\ee
where $m$ and $p^{\mu}$ are the electron mass and momentum, $c=\delta_e-12\sin^2\theta_W$,
$G=\frac{G_F}{\sqrt{2}}$, and $G_F$ is the Fermi constant.
For the speed of relativistic neutrinos we have  $\beta_{(\nu)}^{\mu} \simeq (1,0,0,1)$, thus the effective neutrino potential is $f^{\mu}=G(n,0,0,n)$.
We suppose here that the neutrino flux propagates along the direction of $z$ axis.

\section{Exact solution}
Equation (\ref{modified Dirac equation}) can be solved exactly (see \cite{Balantsev:2014sle}) and for the electron energy spectrum we get
\be\label{branch E_s}
E^{\varepsilon}_{s}(\bs{p})=\varepsilon\sqrt{m^2+\bs{p}_\bot^2+\Big(p_3+A\Big)^2}-A,
\ee
where $A=\frac{Gn}{2}\big(c-s\delta\big)$, $\delta=|\delta_e|$,
$p_3$ is the electron momentum in the direction of the neutrino flux propagation and
$\bs{p}=(\bs{p}_\bot, p_3)$ is the total electron momentum.

Comparing (\ref{branch E_s}) with corresponding spectra of
a \mbox{neutrino \cite{Studenikin:2004dx,Grigorev:2005sw}} or an electron \cite{Grigoriev:2006rv} in nonmoving matter
we conclude that the number $s=\pm 1$ distinguishes two possible electron spin states.

\begin{center}
\begin{figure}[h]
\center{\includegraphics[width=0.8\linewidth]{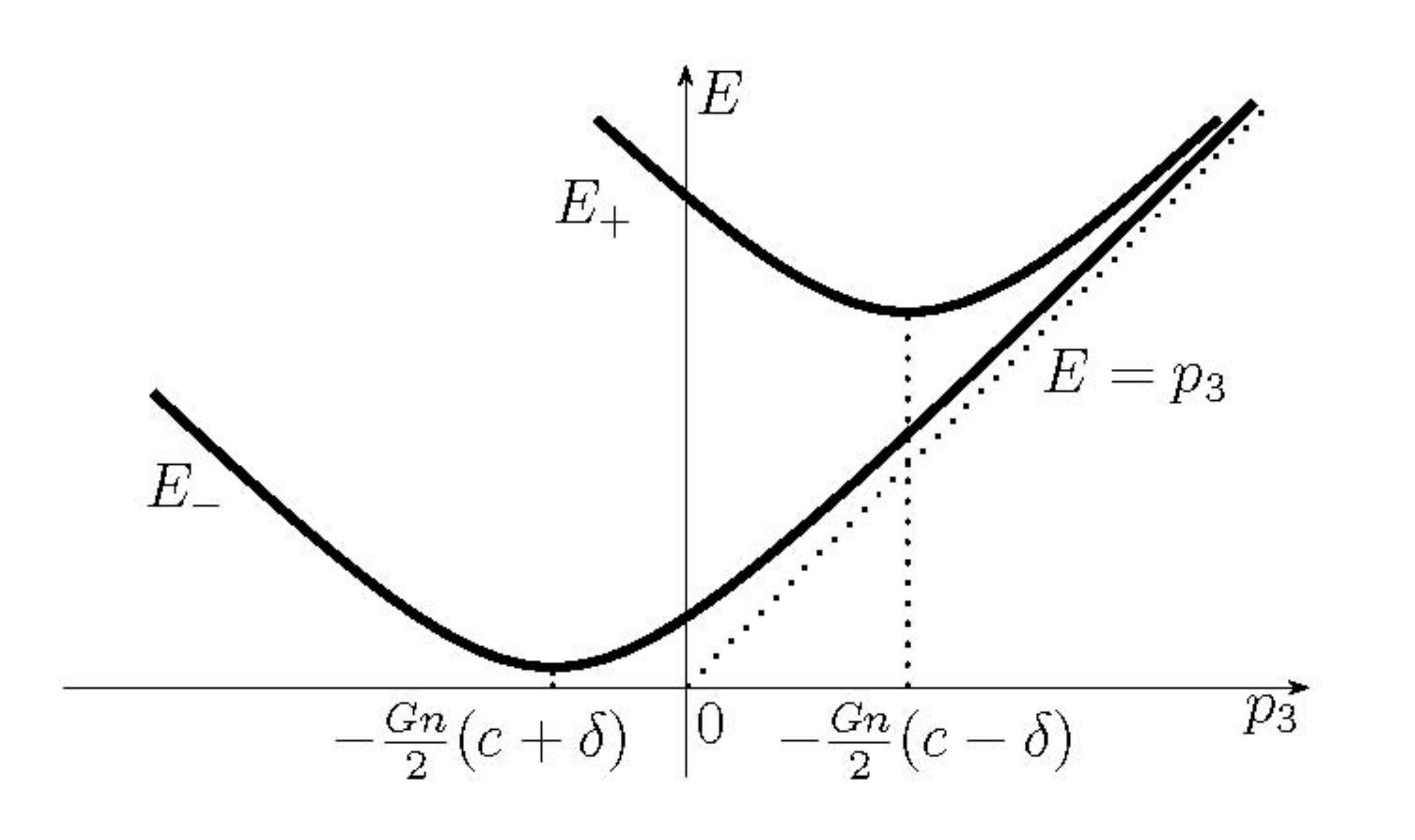}}
\vspace{-0.5cm}
\caption{The dependence of the electron energies in two different spin states, $E_+(\bs{p})$ and $E_-(\bs{p})$,
on the momentum component $p_{3}$.}
\label{figure spectrum}
\end{figure}
\end{center}

Two particular electron energy branches $E^{\varepsilon}_{s}(\bs{p})_{|\varepsilon=+1}=
E_{s}(\bs{p})$ with $s=\pm1$ as functions of the momentum $\bs{p}$ are plotted in Fig.\ref{figure spectrum}.
It is possible to show \cite{Balantsev:2014sle}  that $E_{+}(\bs{p})> E_{-}(\bs{p})$ for any $\bs{p}$.

The exact solution of equation (\ref{modified Dirac equation}) is given by \cite{Balantsev:2014sle} \be\label{psi_i}\psi_i(\bs{r},t)=\rme^{i(-E_+t+ \bs{p}\bs{r})}\tilde{\psi}_i,\ee
\be\label{psi_j}\psi_f(\bs{r},t)=\rme^{i(-E_+t+ \bs{p}\bs{r})}\tilde{\psi}_f,\ee
where
\be\label{initial_and_final_states}
\tilde{\psi}_i=\frac{1}{L^{\frac{3}{2}}C_+}
\begin{pmatrix}
0\\m\\p_\bot\rme^{-i\phi}\\E_{+} -p_3
\end{pmatrix},
\quad
\tilde{\psi}_f=\frac{1}{L^{\frac{3}{2}}C_-}
\begin{pmatrix}
E_{-}-p_3\\-p_\bot e^{i\phi}\\m\\0
\end{pmatrix},
\ee
here $L$ is the normalization length and
\be\label{C}
C_{\pm}=\sqrt{m^2+p_\bot^2+(E_{\pm}-p_3)^2}
\ee
are the normalization coefficients.

\section{Spin light of relativistic electron in dense neutrino flux}

Consider the quantum transition of an electron from one quantum spin state to another with emission of a photon
when the electron moves rapidly towards the dense neutrino flux.
The element of $S$-matrix defining the process amplitude is given by (see \cite{Studenikin:2008qk,
Grigoriev:2006rv}):
\be\label{amplitude basic}
S^{(\lambda)}_{fi} = -e\sqrt{4\pi}\int\,d^4x \bar{\psi}_f(x)({\bs\gamma} \bs{e}^{(\lambda)}{}^\ast)\frac{\rme^{\rmi kx}}{\sqrt{2\omega L^3}}\psi_i(x),
\ee
where $e$ is the electron charge, $\psi_i(x)$ and  $\psi_f(x)$ are the wave functions of the initial and final
electron states in the background neutrino flux given by (\ref{psi_i}) and (\ref{psi_j}), $k = (\omega, \bs{k})$ and $\bs{e}^{(\lambda)}$ ($\lambda=1,2$) are the momentum and polarization vectors of the emitted photon.

\vspace{-0.5cm}

The rest frame in moving background is defined as one where the electron energy $E_+$ gets its minimum,
$\frac{\partial E_+}{\partial\bs{p}}=0$ (see \cite{Studenikin:2004dx,Chang:1988prd,Pantaleone:1991plb,Pantaleone:1992prd}):
$p_3 = -\frac{Gn}{2}(c-\delta)$, $\bs{p}_\bot = 0$.
Thus in general case the initial value of the electron momentum third component  can be represent as
\be
p_3 = -\frac{Gn}{2}(c-\delta) + \tp,
\ee
where $\tp$ is an ``access" of the momentum component over its (minimum) value in the rest frame.
Now we consider the relativistic electrons characterized by the following conditions,
\be
\qquad \tpm \gg m, \quad \tpm Gn\delta \ll m^2,\ \  \text{and} \ \quad \tp<0.
\ee
As for the supernova environment $\frac{Gn}{m}\sim 10^{-8}$, the electron momentum in this case
should be within the range $1\ll\frac{\tpm}{m}\ll 10^8$.

From the energy-momentum conservation law we obtain the expression for the emitted photon energy
\be\label{frequency}
\omega =
\frac{2Gn\delta}{1+\cos\theta+\frac{1}{2}\frac{m^2}{\tp^2}},
\ee
where $\theta$ is the angle between the direction of the $SLe_{\nu}$ and neutrino flux propagation.

It is interesting to compare the emitted photon energies in the case of the considered here  $SLe_{\nu}$ by relativistic electrons and one produced by electrons at rest (see \cite{Balantsev:2014sle}). Taking into account that in the case of nonmoving electrons $\omega=Gn\delta$, for the photons energy ratio (in the case of electron motion against the neutrino flux propagation, $\theta=\pi$) we get
\be
\frac{\omega(\tpm \gg m)}{\omega(\tpm \ll m)}=4\frac{\tp^2}{m^2}\gg 1.
\ee
It follows that there is a reasonable increase of the emitted photon energy in case of the relativistic motion of the  emitters (the electrons).

Using expressions for the amplitude (\ref{amplitude basic}) and the wave functions of the
initial and final electrons (\ref{initial_and_final_states}), and also for the emitted photon
energy (\ref{frequency}) we get
for the $SLe_{\nu}$ total rate and power,
\be
\Gamma=
\frac{16}{3}e^2ma
^3\Big(\frac{\tpm}{m}\Big)^2,\quad
I=
16e^2m^2a
^4\Big(\frac{\tpm}{m}\Big)^4,
\label{radiation rate and power}
\ee
where $a=\frac{Gn\delta}{m}$.
Comparing these expressions with the corresponding characteristics of the $SLe_{\nu}$ in case of nonmoving electrons \cite{Balantsev:2014sle}, we get that
\begin{equation*}
\frac{\Gamma(\tpm \gg m)}{\Gamma(\tpm \ll m)}=4\Big(\frac{\tpm}{m}\Big)^2,\quad
\frac{I(\tpm \gg m)}{I(\tpm \ll m)}=12\Big(\frac{\tpm}{m}\Big)^4.
\end{equation*}
For the case of relativistic electrons $\frac{\tpm}{m}\gg 1$. Thus we show that there should be a reasonable  amplification of the $SLe_{\nu}$ rate and power in case of relativistic electrons.


\section{Effect of plasma}

The electromagnetic wave propagation in the background environment is influenced by the plasma effects.
For the $SL\nu$ in matter these effects have been discussed in details in \cite{Grigorev:2005sw,Kuznetsov:2007ar,Grigoriev:2012pw}.
In \cite{Balantsev:2014sle} we have shown that the effect of nonzero emitted photon mass (the plasmon mass $m_\gamma$) in the case of $SLe_{\nu}$
is not important, $\frac{m_\gamma}{Gn\delta}\ll 1$. The Debye screening of electromagnetic waves (another possible plasma effect) could be important for the $SLe_{\nu}$ radiation propagation if electron number density $N_e<10^{35}\,cm^{-3}$. However, the electron matter with $N_e\sim 10^{19} \ cm^{-3}$ considered here is quite transparent for the $SLe_{\nu}$.

\section*{Conclusions and indications for possible phenomenology}

Let us apply the considered $SLe_{\nu}$ of relativistic electrons in dense flux of neutrinos to an environment peculiar to the supernova phenomena. On the basis of \cite{Kachelriess:2008arXiv} one can estimate the effective neutrino matter density to be $n\sim 10^{35}\,cm^{-3}$, thus the characteristic parameter  $\frac{Gn}{m}\sim 10^{-8}$. As it is discussed in \cite{Liebendorfer:2004aj,Janka:2006fh}, the surrounding interstellar medium
can contain regions with reasonably high electron density relativistically moving
towards the neutrino flux. Under these conditions, the spin light can be emitted by relativistic electrons in the quantum transition from the energy states $E_+$ to the states $E_-$.


From (\ref{frequency}) and (\ref{radiation rate and power}) for the relativistic electrons characterized by  $\frac{\tpm}{m}=10^7$ we get the following estimations for the $SLe_{\nu}$ photon energy, rate and power, respectively,
\be\label{rate_power}
\omega\sim 10^{14} \ eV, \quad \Gamma\sim 10^{10}\,s^{-1},\quad I\sim 10^{21} {eV} {s^{-1}}.
\ee
 The electron number density at the distance $R=10 \ km$ from the star center can be of order
$N_e\sim 10^{19} \ cm^{-3}$. Thus, the amount of $SLe_{\nu}$ flashes per second from $1 \ cm^3$
of the electron matter under the influence of a dense neutrino flux is $N\sim 10^{28} \ cm^{-3} \ s^{-1}$.
For the energy release of $1 \ cm^3$ per one second we get
\be\label{delta_E}
\frac{\delta E}{\delta t \delta V} = I N_e \sim 10^{40}\ eV \ cm^{-3} \ s^{-1}.
\ee

Now let us also estimate the efficiency of the energy transfer from the total neutrino flux to the electromagnetic radiation
due to the proposed $SLe_{\nu}$ mechanism. The total neutrino energy in the neutrino flux
(characterized by $n\sim 10^{35}\,cm^{-3}$ and $\langle E\rangle\sim 10^7 \ eV$) is
\be\label{delta_neutrino_E}
\frac{\delta E_\nu}{\delta V} \sim \langle E\rangle n \sim 10^{42}\ eV \ cm^{-3}.
\ee

It follows that each second a considerable part of neutrino flux energy transforms into
gamma-rays by the $SLe_{\nu}$ mechanism.
The performed studies illustrates an increase of the  efficiency of such energy transfer mechanism
in the case when the emitting electrons are moving with relativistic speed against the neutrino flux propagation
in comparison with the case of nonmoving initial electrons.
We predict that this may have important consequences in astrophysics and for the supernova process in particular.

\section*{Acknowledgments}

We are thankful to Alexander Grigoriev, Alexey Lokhov and Alexei Ternov for many
fruitful discussions on the considered phenomenon. One of the authors (A.S.) is thankful to Arcadi Santamar\'{\i}a, Salvador Mart\'{\i} and Juan Fuster for the kind invitation to participate at the ICHEP 2014 conference and to all of the organizers for their hospitality in Valencia. This study has been partially supported by the Russian Foundation for Basic Research (grant No. 14-22-03043-ofi).











\def\Journal#1#2#3#4{{#1} {\bf #2}, #3 (#4)}

\def\NCA{\em Nuovo Cimento}
\def\NIM{\em Nucl. Instrum. Methods}
\def\NIMA{{\em Nucl. Instrum. Methods} A}
\def\NPB{{\em Nucl. Phys.} B}
\def\PLB{{\em Phys. Lett.}  B}
\def\PRL{\em Phys. Rev. Lett.}
\def\PRD{{\em Phys. Rev.} D}
\def\ZPC{{\em Z. Phys.} C}
\def\JPA{{\em J. Phys.} A}

\end{document}